\documentclass[final]
  {aipproc}

\layoutstyle{8x11single}

\begin{document}

\title{Dark Matter and Galaxy Formation: Challenges for the Next Decade}

\author{Joseph Silk}
{address={ Department of Physics,  University of Oxford,
Denys Wilkinson Building, Keble Road, Oxford OX1 4LN}}

\begin{abstract}
 The origin of the galaxies represents an important focus of current
 cosmological research, both observational and theoretical. Its
 resolution involves a comprehensive understanding of star formation,
 galaxy dynamics, the cosmology of the very early universe, and the
 nature of the dark matter.
In this review, I will focus on those aspects of dark matter 
that are relevant for understanding galaxy formation, and describe the 
outlook for detecting the most elusive  component, non-baryonic dark matter.
\end{abstract}

\maketitle

\section{Introduction}

Dark matter and galaxy formation are intimately related.  This applies
equally to baryonic and to nonbaryonic dark matter.  In this talk, I will
review the global budget for baryons and discuss the issue of dark baryons.
I will describe the role of nonbaryonic dark matter in galaxy formation, and
give an overview of the prospects for detection of cold dark
matter.

A confluence of data on the cosmic microwave background
temperature fluctuations, large-scale galaxy redshift surveys, quasar
absorption line structure of the intergalactic medium, and distant supernovae
of Type Ia have led to unprecedented precision in specifying the cosmological
parameters, including the matter and energy content of the universe.  The
universe is spatially flat, $\Omega = 1.02 \pm 0.02$, and dominated by dark
energy $\Omega _{\lambda} = 0.70 \pm 0.3$ with equation of state $w \equiv
\frac{\rho}{\rho c^{2}} = -1.02 \pm 0.16,$ nonbaryonic dark matter amounting
to $\Omega _{m} = 0.27 \pm 0.07$, and the baryon content $= 0.0044 + 0.004$.
The latter number incorporates a value of the Hubble content $H_{0} = 72 \pm
5 \rm km s^{-1} Mpc^{-1}$. 

A  major assumption underlying the quoted
errors is the adoption of priors. In particular, primordial gaussian
adiabatic, scale-invariant density fluctuations are adopted.  If, for example,
an admixture of 30 per cent isocurvature fluctuations is included, 
consistency with CMB data is still obtained but
the error
bars are expanded by up to an order of magnitude \cite{buch04}.  
Another assumption is that
the fine-structure constant is actually constant.  Allowing this to vary also gives
further freedom, especially in the baryon density.  

A strong case
for the dominance of dark matter in galaxy clusters was made as long ago as
1933.  It is  remarkable that our understanding of its nature has not advanced
since then.  Of course, modern  observations have led to an increasingly
 sophisticated exploration of the distribution of dark matter,
now confirmed to be a dominant component relative to baryonic matter over
scales ranging  from those of galaxy halos to that of the particle horizon.

\section{2. Global Baryon Inventory}

There are three methods for determining the baryon fraction in the high
redshift universe.  The traditional approach is via primordial
nucleosynthesis of $^{4}$He, $^{2}$H and $^{7}$Li.  The primary
uncertainties lie in the systematic errors associated with ionisation corrections for
$^{4}$He, and extrapolation to primordial values via corrections for
synthesis of $^{4}$He and destruction of $^{2}$H and $^{7}$Li in stars.  A
unique value of $\Omega _{b} = 0.04 \pm 0.02$ is generally consistent with recent data,
although there is some tension between $^{2}$H, on the one hand, which in
principle is the most sensitive baryometer and favours a higher $\Omega
_{b},$ and both $^{4}$He and $^{7}$Li.  
This tension has recently been
increased \cite{lamb04}
by the demonstration of a [Li/Fe] gradient of $\sim \frac{\rm 0.1
dex}{\rm dex}$ in extremely metal-poor halo stars with
 $\left[\frac{Fe}{H}\right]<-2,$ indicative of a role for pregalactic stellar destruction 
of primordial Li,  as well as by  determinations of
$\frac{^{6}Li}{^{7}Li} = 0.05-0.08$ that indicate a 10-15\% spallation 
contribution  to
$^{7}$Li in this metallicity range.  Hence more generous error bars may be
preferred, at least until the role of systematic effects 
such as atmospheric depletion of depletion
are fully understood.

 A completely independent probe of $\Omega _{b}$ comes
from measuring the relative heights of the first 3 peaks in the acoustic
temperature fluctuations of the cosmic microwave background.  With the
conventional priors, the data yields excellent agreement between the baryon
abundance at $z \approx 1000$ and $z\approx 10^{9}$.  Relaxation of the
priors increases the error bars, but the central value is relatively
robust. Yet another  independent measure of $\Omega _{b}$, this time at
$z \approx 3,$  comes from modelling the Lyman alpha forest of the
intergalactic medium.  This depends on the square root of the ionizing photon
flux, in this redshift range due predominantly to quasars.  The inferred value of $\Omega _{b}$ 
is again 0.04, with
an uncertainty of perhaps 50\%.
Finally at $z \sim 0$, one only
has a reliable measure of the primordial baryon fraction in galaxy clusters, which may
be considered to be laboratories that have retained their primordial baryon
fraction.  The observed baryon fraction in massive clusters is about 15\%,
which is consistent with $\Omega _{b} = 0.04$ for $\Omega _{m} = 0.28$, the
WMAP-preferred value.

 Let us now evaluate the baryon fraction at
the present epoch, both on galactic scales and in the general field
environment. The following is 
an updated summary  of the baryon budget recently presented by 
Fukugita and Peebles \cite{fuku04}.

Stars in galactic spheroids account for about twice
as much baryonic mass as do stars in disks.  Disks dominate the (blue) light
but spheroids have higher mass-to-light ratios.  The total stellar
contribution is about 15\% of the total baryonic abundance of 0.04.  Rich
clusters only account for 5\% of the galaxies in the universe, and so all of
the hot diffuse gas in clusters, which account for 90\% of cluster baryons,
only accounts for about 5\% of the total baryonic budget.

 Cold
intergalactic gas at the current epoch is mapped out in Lyman alpha
absorption towards quasars. Identified with the Lyman alpha forest observed at high redshift,
the low redshift counterpart  is sparser. Its detection is more difficult,
requiring a UV telescope such as HST or FUSE.  However it is found to dominate the known baryon fraction
today, and amounts to about 30\% of the total baryon fraction \cite{stoc04}

In summary, some fifty percent of the baryons in the local universe have been detected
and mapped.  There are indications, motivated as much by theory
as by observations at this stage, that the remaining baryons are in the
warm intergalactic medium (WIM)  at a temperature of $10^{5} -
10^{6}$K.  Simulations of structure formation indicate that some
intergalactic gas is shocked to a temperature of $10^{5} - 10^{6}$K.
Much of this gas has not yet fallen into galaxies.  According to the
simulations, up to about 30\% of the baryons are heated by the present
epoch and remain diffuse.  This fraction is an upper limit because the
simulations lack adequate resolution, and moreover the amount of shock-heated gas is 
controversial \cite{birn03}. Even more significantly, the theory of galaxy
formation, as currently formulated, predicts that the WIM is
metal-poor, in that those galaxies where most of the stellar mass
resides, namely the massive galaxies, are energetically
incapable of ejecting very much in the way of metal-enriched debris \cite{spri03}.

 However,
observations are confirming the existence of  some
WIM, in particular via detection of redshifted rest-frame UV  OVI absorption towards quasars,
extended soft x-ray emission near clusters
\cite {zapp04}, and OVII/OVIII x-ray absorption along
lines of sight to AGN.  
The oxygen abundance exceeeds [O/H]> -1.5  at $z\sim 2.5$
\cite{simc04}.
In practice,  too few lines of sight have so far been probed to say a great deal
about the WIM mass fraction.

In summary, something
like 80 percent of the baryons at present have either been detected or are
plausibly present with detection being imminent.  One could conclude
\begin{center}$\Omega _{b, observed} = 0.032 \pm 0.005.$\end{center}
Clearly the case for 10 - 20\% of the local baryons being unaccounted for and
dark is possible but far from  convincing given the WIM uncertainties.
If the WIM is indeed the dominant gas reservoir, there are strong implications for feedback
from galaxy formation, in order to account for the observed enrichment of the WIM.
Strong enrichment is indeed found for the intracluster medium, and this most likely is a consequence of 
early galaxy outflows. However the generation of these outflows is not understood.

One clearly needs to establish a more convincing case for the WIM before pursuing the impact of massive
gas  outflows on the early evolution of 
 the typical field galaxy. Nevertheless,
 since the  possible mass in unaccounted-for dark baryons is on the order of the
baryon mass in stars, it is clear that such a result would profoundly affect our theories of galaxy formation
and evolution. Hence demonstrating that these baryons are not present in the Milky Way is a useful exercise.

\section{3. Confirmation of baryonic shortfall}

A detailed census of both the Milky Way and M31 confirms the lack of baryons
in the amount predicted by primordial nucleosynthesis.  The virial  mass measured
dynamically for the Milky Way from the HI rotation curve, dwarf galaxy
orbits, and globular cluster peculiar velocities, amounts to $\sim 10^{12}M
_{\odot}$.  This is valid to a galactocentric radius of 100kpc. The baryon mass, including stars and
gas is $(6-8) \times 10^{10}M_\odot$.  However, the expected baryon fraction,
both as observed at high redshift and in galaxy clusters, and especially as
inferred from primordial nucleosynthesis and the CMB data,
 is about 17\%.
This is the initial baryon fraction when the Milky Way formed.  A similar
shortfall, amounting to a factor of about 2, is found for M31.

There are two possibilities for the "missing" baryons.  Either they are
present in the galaxy halo and as yet undetected, or they have been ejected
via energetic outflows early in the history of the galaxy.  
Intensive
searches for compact halo objects have been performed via gravitational
microlensing of several million stars in the Magellanic Clouds.  The EROS and
MACHO experiments set the following limits,
for more than 5 years of data: no more than 20 percent of the
dark halo mass can be in objects in the mass range $\sim 10^{-8}\rm M_\odot$ to $\sim 10\rm M_\odot$,
with a detection claimed by the MACHO experiment that saturates this limit for objects of mass $\sim$
0.5M$_\odot.$ \cite{alco00,afon03}
 
The most plausible candidate for MACHOs of this mass are old
halo white dwarfs.  This requires a stellar initial mass function for the
protogalaxy that forms the first stars with high efficiency in a narrow mass
range $(4-8M_\odot)$.  While this seems implausible, it cannot be ruled out by
theoretical arguments, one possible signature being that of occasional Type
Ia supernovae.  However old white dwarfs are still emitting light, albeit
weakly, at visible wavelengths, and proper motion searches for faint
candidates have imposed strong limits on the halo white dwarf mass fraction
of between $\sim 2$\% relative to the local dark matter density \cite{crez04}
and  $\sim 0.2$\% \cite{spag04}.
It seems reasonable to conclude that halo white dwarfs cannot
account for more than a quarter of the unacounted-for baryons, and this is most
likely an overestimate.

One can imagine even less credible initial  mass
functions that would allow, say, ten percent of the dark halo to consist of
primordial brown dwarfs, low mass primordial black holes, or even compact
dense clouds of cold molecular gas.  All of these possibilities have been
studied as possible explanations for halo dark matter. Even if one's goal is
only to account for halo baryonic dark matter, requiring even
$10^{10}M_\odot$ to be in such a form stretches astrophysical credibility.
But this cannot be ruled out.

 A more plausible  direction for investigation is that
the "missing" baryons have been ejected from the galaxy, in the form of a
vigorous, early galactic wind.  Such a wind, if it occurs presently, could
involve very little mass outflow. Observations indicate that at the present epoch, vigorous
winds are exceedingly rare, and are seen only in low mass, star-bursting galaxies.
 In the early galaxy, however, the star
formation rate was much higher, and the situation could have been quite
different with regards to mass loss.  Evidence for early winds comes
indirectly from the highly enriched intracluster medium, whose mass exceeds that in 
the stellar component of cluster galaxies by a factor of several.  The substantial amount of
metals in the intracluster gas, and even the presence of magnetic fields, are
most likely accounted for via ejection in early galactic winds.  

At high redshift, the
substantial population of the Lyman break galaxies (LBG)  at $z \sim 3-4$ show
broad linewidths displaced systematically to the blue
by several hundred kilometres per second
  for  the interstellar gas relative
to the absorption
lines of the stars \cite{stei04}.
Moreover, stacked spectral energy distributions of LBGs
seen in projection near background quasars show evidence of a proximity
effect, with a $\sim 1 \rm \,Mpc$ hole (comoving) inferred from  the lack of Ly$\alpha$ and CIV
absorption \cite{adel03}.  An energetic wind from galaxies with stellar mass similar to that
of the Milky Way is inferred to have occurred,
or at least, to provide the simplest explanation of these observations.
Some of these galaxies most likely are massive, as their spatial clustering
strongly favours their being the precursors of  low redshift ellipticals \cite{adel04}.

The principal
counterargument comes from wind simulations.  While it is unanimously agreed
that dwarf galaxies, with masses below $10^{7} - 10^{8}M_\odot$, and escape
velocities below 50kms$^{-1}$, are easily stripped of gas by
supernova-driven winds, problems arise in driving winds from more massive
galaxies.  For disk galaxies, it is found that even for galaxies of mass
10$^{9}$-10$^{10}$M$_\odot$, the supernovae ejecta stream out  in a hot wind
but most of the interstellar gas remains in the disk \cite{macl99}.

For forming galaxies,
when the gas is more spherically distributed, ejection in a wind becomes
inefficient for masses above about 10$^{10}$M$_\odot$, according to the most
recent multi-phase interstellar medium simulations \cite{spri03}.  These simulations adopt
current supernovae rates and  energetics per unit baryonic mass, along with a solar neighbourhoopd initial mass function,
that is to say a rate of 
type II supernovae of 10$^{51}$  energy input per 200M$_\odot$ of gas that forms
stars.  This rate assumes  a local  fit to the initial mass function \cite{krou02}.

 However
in addition to the observational indications, semi-analytical galaxy
formation theory requires a wind to have ejected approximately half of the
baryons from even the most massive galaxies.  Otherwise, one finds that almost
all
of the gas that can cool within a Hubble time does cool and form stars, and the predicted
luminosity function strongly disagrees with observations for luminosities
above 2-3 times the galaxy 
characteristic luminosity, $L_{\ast}\sim 10^{10}\rm M_\odot$ \cite{bens03}.
Related model malfunctions  include unacceptably  recent and inefficient star formation for distant
massive galaxies as studied in deep surveys \cite{thom04}.

\section{4. What could be wrong with the simulations?}

The numerical simulations of galactic outflows must cope with a variety of
hydrodynamical and gravitational processes, including star formation,
supernovae explosions, gas heating and cooling in a multi-phase interstellar
medium, and gas escape from the galactic gravitational field.  Hitherto, it
has been necessary to severely approximate much of the relevant physics.  For
massive galaxies, winds are suppressed  as the outflowing heated gas runs into
surrounding, cold infalling gas, and most of the energy input is radiated
away. Only about 2 percent of the initial supernovae energy is useful for
expelling gas.  

The situation may not be as bleak as depicted by the
simulations.  One omission due to lack of resolution is the effect of both
Rayleigh-Taylor and Kelvin-Helmholtz instabilities.  The former help the hot
medium break out of the galaxy and enhance the wind efficiency.  The latter
enhances  entrainment  of the cold gas into the hot supernova-heated medium
and can help account for the observed enrichment of the intergalactic medium.  
While
the situation with regard to outflows may be alleviated 
in this fashion 
for low and intermediate mass galaxies, more drastic
measures are required for massive galaxies.  These may include any of the
following: use of a top-heavy initial stellar function to enhance the
supernova rate or appeal to an increased frequency of hpernovae at early
epochs relative to supernovae, or finally,  recourse to outflows from active
galactic nuclei.  Any or all of these may occur.

If indeed
substantial mass loss via a wind occurs, then a plausible ansatz is that
$\dot{M}_{outflow}\sim\dot{M}_{\ast}$ as observed in nearby starbursts,
where the mass injection rate into the hot x-ray emitting diffuse gas
is comparable to the star formation rate \cite{summ04}.
  This
means that about as much gas is ejected as is retained in stars.  Such a
conclusion is consistent with the observed baryon fraction in the Milky Way
and M31, the two best-studied moderately massive galaxies.  One can also
understand the heavy element abundance observed both in the intercluster
medium and in the warm ($T\sim10^{6}$K) intergalactic medium detected in OVI
absorption.  While the baryon fraction is probably not a major problem for
consensus cosmology, I now turn to the issue of cold dark matter, and its
relation to structure formation.

\section{5. Galaxy formation and CDM: the good, the bad and the ugly}

There are some noteworthy success stories for cold dark matter (CDM).  First and
foremost is its success in predicting the initial candidates for structure
formation that culminated in the discovery of the cosmic microwave background
temperature fluctuations. The amplitude of the Sachs-Wolfe effect was
predicted to within a factor of 2, under the assumption, inspired qualitatively by
inflation, but quantitatively by the theory of structure formation via gravitational instability
in the expanding universe, of adiabatic scale-invariant initial density fluctuations.  
A 
direct confrontation with this theory was first  met  with the
detection and mapping of the acoustic peaks.  These are the hallmarks of
galaxy formation, first predicted some  three decades previously, and
demonstrate the imprint of the density fluctuation initial conditions on the
last scattering surface of the CMB at $z\sim 1000$.

 Another
dramatic demonstration of the essential validity of CDM has come from the
simulations of the large-scale structure of the universe.  The initial
conditions, including gaussianity,  are specified,  growth occurs  by gravitational instability, and the
sole requirements on dark matter are that it be weakly interacting and cold.
Thus was born CDM, and the CDM scenario works so well that we cannot
easily distinguish the artificial universe from the actual universe mapped via
redshift surveys.  More to the point, perhaps, is that the simulations are
used to generate mock galaxy catalogues and maps that yield precise values of
the cosmological parameters, in combination with the CMB maps.

Dark matter-dominated halos of galaxies are another generic success of
CDM, as mapped out by rotation curves.  However the detailed predicted
properties of halos do not seem to be well-matched to observations.
There is considerable scatter in the predictions of high resolution
simulations for the structure of galaxy halos.  Nevertheless, the
predicted dark matter cusps $(\rho\propto r^{-\alpha}$ with $1 <\alpha
<1.5)$ are not found in most low surface brightness dwarfs, nor is the
predicted dark matter concentration ($C\equiv r_{200}/r_{s}\sim 5-10$,
where $r_{200}$ is the radius at density contrast 200 and $r_{s}$ is
the halo scale length) consistent with the dark matter distribution in
barred galaxies, possibly including our own galaxy, nor finally is the
predicted number of satellites similar to the observed satellite
frequency.  
In general, many  observed halos seem to have softer cores,
lower concentrations, and less clumpiness than predicted by the
simulations. 

However it has been argued that inclination, triaxiality  and non-circular orbits
make the dwarf situation unclear \cite{haya04}, quite apart from the fact that
dwarf galaxy formation is not understood. Of course the same may be said for
bars.  The situation for early-type galaxies is at least as
controversial.  Indeed, for very round ellipticals, at least in
projection, in low density environments and not especially luminous,
studies of the distribution and kinematics of
planetary nebulae suggest that mass traces light to $\sim5$ effective
radii \cite{roma03}.  However, the opposite conclusion is inferred for massive
early-type galaxies, which display evidence for as much as a 50\%
contribution of dark matter within  $\sim1$
effective radius \cite{treu04}.

All of these issues have been debated. For example,  reformation of bars by gas
infall can avoid the problem of bar spin-down by dynamical friction,
and astrophysical processes, discussed below,  can render the dwarf satellites
optically invisible.  Hence it is difficult to be definitive about any
possible contradiction between theory and observation.
Certainly, on the baryonic
front, the most accepted problem is the loss of angular momentum by the
contracting and cooling baryons in the dark halo.  The resulting disks are
far too small. 
These various difficulties for galaxy
formation theory have stimulated a variety of responses.

\subsection{Resurrection via modifying fundamental physics}

Suppose that one changes the nature of the dark matter.  Increasing the
scattering cross-section helps alleviate several of the problems, such as
cuspiness  and clumpiness.  However the resulting dark halos are too spherical.
Another approach modifies the law of gravity.  Indeed, one may be able to
dispense entirely with dark matter.
These approaches seem rather drastic,
however, and I believe that one should argue that all alternatives should be fully explored
before tinkering  with fundamental physics.

\subsection{Resurrection via astrophysics}

The obvious addition is stellar feedback.  This can heat the baryons, and
help reduce the loss of angular momentum.    If the feedback is strong,
mass loss is a likely outcome.
The observed baryon
fraction and  the galaxy luminosity function for the most luminous
galaxies  both point to a possible loss of  half the baryons during the
galaxy formation process \cite{silk03}. However
 to eject up to half the baryons
may require more than normal stellar feedback, at least for galaxies
comparable to, or more massive than, the Milky Way.  One can appeal to a
top-heavy IMF that would yield up to an order-of-magnitude more supernovae
per unit mass of baryons, to an augmented  fraction of hypernovae relative to
supernovae, or to outflow generated by Eddington luminosity-limited accretion
onto a supermassive black hole.  Outflows may also be effective at reducing
the dark matter concentration, at least for dwarfs \cite{read04}.
  
Production of a soft core is best  achieved for a massive galaxy by
dynamical heating, as has been studied for the case of a rapidly rotating
central baryonic bar \cite{holl03}, although a contrary view is expressed in  
\cite{sell03}. 
Such bars are likely to be generic to galaxy formation
via mergers, and if gaseous would leave little in the way of stellar
tracers.
Dynamical feedback also occurs via tidal evolution, and this can 
account for both the frequence and distribution of dwarf galaxies \cite{krav}.

\section{6. Observing cold dark matter}

The best way forward is to directly measure the halo properties by observing
cold dark matter directly or indirectly. Direct detection is
sensitive both to the local density of CDM and to its local phase  space
density.  There is a candidate, motivated by supersymmetry, the LSP, usually
considered to be massive with $m_{x}\sim 100$GeV,
  the SUSY breaking scale,
and 
generically 
known as the neutralino or WIMP.  However light LSPs, such as the axino, are also possible,
and there is even a LSP with purely gravitational interactions, the
gravitino.  However, in general, the WIMP undergoes elastic interactions with
ordinary matter and is therefore potentially detectable via laboratory
experiments.  Early universe freezeout yields a mass estimate; more
specifically, the annihilation cross-section is inferred to be of order
$\Omega^{-1}_{\chi}\sigma _{weak}$, and depends, via SUSY, on the WIMP mass.
The corresponding elastic cross-section is model-dependent, but most models
spans the range 10$^{-10}$ to 10$^{-6}$pb for a relic abundance $\Omega
_{x}h^{2}\approx 0.1$.

\subsection{Direct detection}

Scattering of WIMP particles leads to nuclear recoils that can be measured by
three different techniques: scintillation, phonon production, and ionization.
The various experiments currently underway use different combinations of
these techniques.  Only one experiment, now running for 7 years, has reported
a positive result, using NaI scintillation and a claimed detection of annual
modulation, to yield a model-dependent detection of $m _{\chi} = 50(\pm
10)\rm GeV$ with a
cross-section of $7(\pm 1) \times 10^{-6}$pb.  However other experiments,
including Edelweiss, ZEPLIN and CDMS2, report a lower upper bound in the
cross-section, with the more recent limit being $\sigma _{\chi} < 4 \times
10^{-7}$pb at 60 GeV \cite{cdms04}.

\subsection{Indirect detection}

Annihilations currently occur in the dark halo, although the annihilation
time-scale $\sim (n _{h < \sigma v >_{ann}})^{-1} \sim 10^{26}(
{T_{f}/\rm GeV})^{{3/2}}$s, where $T_f$ is the freeze-out temperature.
  The annihilation products are potentially
observable in the form of high energy $\gamma,e^{+}, \bar{p}$ and $\nu$, and
are enhanced by the effects of halo clumpiness.  There are tentative
indications of possible detections of $e^{+}$ and $\gamma$.  A positron feature
$\frac{e^{+}}{e^{+}+ e^{-}}$ is seen above 10GeV that cannot easily be
attributed to secondary production of $e^{+}$.  A modest clumpiness boost is
required for the measured flux to lie in the range allowed by annihilation
models combined with cosmic ray diffusion \cite{balt02}.
Both the high galactic
latitude gamma ray background and the unresolved diffuse gamma ray flux
towards the galactic centre have relatively hard spectra that seems to be
inconsistent with cosmic ray spallation and the ensuing  $\pi ^{o}$ decays.  One
possible explanation is in terms of population of hitherto unresolved
discrete gamma ray sources, such as blazars in the extragalactic case or low
mass x-ray binaries in the galactic case \cite{ulli02}.  Similar boost factors,
of 10-100,  from
dark matter clumpiness are required to that invoked for positron annihilation, if
both the extragalactic and galactic diffuse gamma ray components have a WIMP
annihilation origin.

\subsection{A radical suggestion}

The Integral SPI detector has measured a substantial diffuse flux of
electron-positron annihilation line emission at 511 keV from throughout the
galactic bulge.  Some $10^{43}$ photons s$^{-1}$ are generated over a
region  that extends up to 3 kpc from the galactic centre.  There is no indication
of any positron annihilation emission from any bulge source, such as might be
connected with decays of Type II supernova-ejected radioactive $^{26}$Al or 
$e^{+}-e^{-}$ jets from x-ray
binaries.  This therefore has led to consideration of CDM annihilation as a possible 
explanation \cite{boeh04}.

 The principal novelty of such a hypothesis arises
with the mass required for the annihilating particle.  It must have a mass of
$\sim$10 MeV, as a much heavier particle would annihilate via pion production
and produce an excessive flux of diffuse gamma rays from $\pi ^{o}$ decays.
From the measured flux and angular distribution, one immediately infers the
required cross-section and radial profile, namely $\sigma _{ann}\sim
10^{-5}$pb and  $\rho_{\chi}\propto r^{-1/2}$.  The profile is close to what
is expected from CDM models,
as  inferred from  rotation curve and microlensing
modelling (actually, the derived CDM profiles  are disputed for the Milky Way but a
profile softer than NFW is inferred for barred galaxies and for LSB
dwarfs). The required cross-section is very low, however, compared
with the freeze-out value at kT$\sim {m_{\chi}/20}$, namely $\sigma
_{ann}\approx ({0.2/\Omega _{\chi}})$pb.  One can reconcile
 the observed low annihilation 
cross-section required for the 511keV flux by assuming that the relativistic
freeze-out  limit is S-wave
suppressed, so that
$\sigma _{ann}\propto
\left(\frac{m^{2}_{\chi}}{m^{4}_{U}}\right)\left(\frac{v}{c}\right)^{2}$\nonumber\\
This naturally reduces the low temperature value of the halo annihilation
cross-section relative to the freeze-out value by a factor $(v/c)^{2}\sim
10^{-5}$.  

There is  a price to pay however for the low mass, namely the
introduction of a new light gauge boson $m_{U}\propto m_{\chi}^{{1/2}}$,
ordinarily  comparable in  mass to the Z boson if $m_{\chi}$ is at the SUSY
breaking scale.  A mediating $m_{U}\sim 0.1-1$GeV could have observable
consequences, for example with regard to the magnetic moment of the muon,
and these are being investigated. 

 One should also eliminate
possible astrophysical sources of the 511 keV line.  The most
promising  of these is
the population of low  mass x-ray binaries, which have a bulge distribution
and are known to occasionally have high energy jets and outflows.  However
there has hitherto been no association of 511 keV emission with any class of
discrete sources.

\subsection{An equally radical suggestion}

Three atmospheric Cerenkov radiation telescopes have recently reported
the detection of TeV photons from the Galactic Centre.  HESS has the
most significant detection.  The supermassive black hole associated
with the SagA* radio source is measured to have a mass of
3$\times$10$^{6}$M$_\odot$, and x-ray measurements indicate a low
accretion rate.  Hence a source of $\gamma$-rays powered by accretion is
unlikely.  One could appeal to a high energy cosmic ray accelerator
associated with the central black hole.  However the low observed
accretion rate may (weakly) argue against this.  An acceleration power
in TeV electrons or EeV protons of 10$^{36}$-10$^{39}$ ergs s$^{-1}$,
respectively, is required, where the bolometric luminosity is only
10$^{36}$ ergs s$^{-1}$ (or $10^{-8}L_{\rm Edd}$).

 An annihilation explanation requires WIMPs of mass at least 10-20
TeV.  In this case, the observed hard spectrum is naturally explained
\cite{berg04, horn04}. However there are difficulties that arise in
reconciling the WMAP-constrained value of $\Omega_\chi$ with the
cross-section required to account for the HESS luminosity of
$10^{35}\rm s^{-1}$ above 200 GeV with half-width of 6 arc-minutes.  To
arrive at the required relic density for a 20 TeV neutralino  mass, one has
to fine-tune the particle physics annihilation channels via
co-annihilations. The $\Omega_\chi$ constraint prefers a  cross-section
around 1 pb. 
The natural value of the cross-section at 20
TeV tends to be
lower than 1 pb, because of the unitarity scaling that sets in at
large masses, and this results in WIMP overproduction:  $\Omega _{\chi}$ is too high.
However, for
a typical NFW profile, the inferred cross-section to account for the
observed gamma ray flux at 10 TeV is about 10pb, and is even larger
for a softer core. In this case,
the inferred relic density is too low, only $\Omega _{\chi}\sim
0.03$.

To reconcile these conflicting requirements is not straightforward.
The simplest  option is to relax the relic density constraint.
Suppose that the 20 TeV WIMPs are subdominant. One can now tolerate a larger cross-section.
Particle physics fine-tuning is required via co-annihilations, but this is rarely an unsurmountable  problem.

Although it appears to be very unnatural that the LSP mass would be
any heavier than a few TeV, with a high degree of fine-tuning,
co-annihilations can allow for much heavier LSPs. Even in this case,
however, it would seem very unlikely that the LSP mass could be any
heavier than 20 TeV, at least in the simplest classes of models.  The
following scenario might then apply.  One would have two types of
stable particle dark matter, as appropriate to N=2 SUSY \cite{faye04}.
The light particle $(m_{\chi}\sim \rm 10MeV)$ would be the principal
dark matter component, and annihilate via $e^{+}e^{-}$ to produce the
511 keV flux.  The subdominant particle, with mass $\sim$10-20 TeV,
would account for the HESS flux.

An alternative is the following. Suppose we settle for the lower cross-section
as inferred from the relic WIMP density. Theory certainly has an easier 
time arriving at this goal.
Then we need to boost the annihilation flux at the centre of the galaxy.
It is unlikely we can appeal to the usual CDM clumpiness boost factor, because  any clumps would be tidally disrupted.
It is then appealing to reconsider
the possibility of a spike
of dark matter around the central SMBH within its zone of influence,  a parsec or so.  This occurs naturally for adiabatic
formation of the SMBH, via the response of the CDM halo,
and yields,  in principle, an observable gamma ray signal from generic
CDM annihilation models \cite{bert02}.  A spike formed in a
pregalactic SMBH would survive infall of the SMBH by dynamical friction to
the centre of the Milky Way galaxy.  This  works best if the SMBH forms by
baryonic accretion rather than by black hole mergers,
although only major mergers are potentially catastrophic for a spike
\cite{ulli01}.  The survival of a spike seems not unlikely
because (a) there is no theoretical understanding of the "final parsec"
problem of merging black holes, (b) minihalo mergers
in hierarchical galaxy formation
yield too  few close-in SMBH  candidates for successful mergers to prevail
in  the final system, and
(c) forming  the very massive SMBHs seen at  $z\geq$ 6 requires an accretion
formation mechanism given the limited time available. The adiabatic spike,
which  has profile $\rho\propto r^{-\gamma}$  with $\gamma>\frac{3}{2}$,
dominates accretion and would yield the HESS point-like source but be
unobservable at INTEGRAL/SPI resolution.

\section{The Future}

Baryon dark matter will most likely be mapped out within five years.  The
intergalactic medium is the major repository where large uncertainty remains.
The warm intergalactic medium can be studied via highly  ionised oxygen, both in UV
absorption and in x-ray emission.  This most likely will require dedicated
experiments that are being planned.  

Of course to distribute the oxygen and
other elements into the WIM/ICM requires a greatly improved understanding of
galactic outflows.  Considerable improvements will be needed in the accuracy
and resolution of simulations of galactic outflows.  Can the escape rate of
gas be of the same order as the star formation rate in massive
young galaxies?  It will require improvement in the input physics of star
formation as well as in the numerical sophistication of the codes before this
question can be fully considered.

Advances on the non-baryonic
matter front seem equally likely.  Of course, here there is a big assumption,
that the elusive dark matter particle is a WIMP.  Were it to be a light
gravitino or  an axion, almost all of the searches would be
frustrated. Nevertheless there are more than a score of dedicated
searches underway for direct and indirect detection of non-baryonic dark
matter.  These include searches for annihilation products, including positrons
and antimatter (PAMELA, AMS2), high energy neutrinos from the sun (ANTARES,
ICECUBE), and gamma rays (GLASST, HESS, VERITAS). It will be
necessary with all of these searches to correlate complementary signals and corroborate astrophysical 
detections with accelerator evidence of existence of the relevant particle.
Such evidence may be beyond the reach of the LHC, but a future linear  collider
should be able to provide the clean signature needed to identify the SUSY
LSP, provided that the WIMP mass is below 1TeV. If the WIMP mass is greater, then ACT (gamma ray
telescopes) may become the unique hope for detection.
Other "smoking guns" include detection of gamma ray line emission and
confirmation of annihilation signals associated with nearby dwarf galaxies
and with the Galactic Centre, where primordial concentrations of dark matter
should exist, by both spectral and spatial resolution.
\begin{theacknowledgments}
  I thank my colleagues, including R. Bandyopadhyay, C. Boehm,
P. Ferreira, D. Hooper, H. Mathis,. J. Taylor and H. Zhao,  for many
discussions of relevant topics. I am also indebted to Professor Piet
van der Kruit for hosting me as Blaauw
Visiting Professor at the Kapteyn Institute in Groningen, where this
review was completed.
\end{theacknowledgments}

\end{document}